# Reevaluation of radiation reaction and consequences for light-matter interactions at the nanoscale

M. Scalora[1], M.A. Vincenti[2], D. de Ceglia[3], N. Akozbek[4], M. J. Bloemer[1], L. Roso[5], J. Trull[6], C. Cojocaru[6], and J. W. Haus[7]

[1] *Charles M. Bowden Research Center, AMRDEC, RDECOM, Redstone Arsenal, AL 35898-5000 - U.S.A.*

[2] *Department of Information Engineering – University of Brescia, Via Branze 38 25123 Brescia, Italy*

[3] *Dipartimento di ingegneria dell'informazione – University of Padova, Via G. Gradenigo, 6/B, Padova, Italy*

[4] *AEgis Technologies Inc., 401 Jan Davis Dr. 35806, Huntsville, AL U.S.A.*

[5] *Centro de Láseres Pulsados (CLPU), Parque Científico, E-37185, Villamayor, Salamanca, Spain*

[6] *Physics Department, Universitat Politècnica de Catalunya, Rambla Sant Nebridi 22, Terrassa, Barcelona, Spain*

[7] *Department of Electro-Optics and Photonics, University of Dayton, Dayton, OH 45469-2951*

michael.scalora.civ@mail.mil

**Abstract**

In the context of electromagnetism and nonlinear optical interactions damping is generally introduced as a phenomenological, viscous term that dissipates energy, proportional to the temporal derivative of the polarization. Here, we follow the radiation reaction method presented in [G. W. Ford and R. F. O'Connell, *Phys. Lett. A,* 157, 217 (1991)], which applies to non-relativistic electrons of finite size, to introduce an explicit reaction force in the Newtonian equation of motion, and derive a hydrodynamic equation that offers new insight on the influence of damping in generic plasmas, metal-based and/or dielectric structures. In these settings, we find new damping-dependent linear and nonlinear source terms that suggest the damping coefficient is proportional to the local charge density, and nonlocal contributions that stem from the spatial derivative of the magnetic field and discuss the conditions that could modify both linear and nonlinear electromagnetic responses.

Historically, the notion of radiation reaction, or radiation damping, was elaborated by Abraham and Lorentz soon after the development of electron theory during the early 1900s [1,2]. A relativistic covariant form of the equations was published by Dirac [3]. The theory accounts for the fact that an accelerated charge simultaneously emits and reacts to its own electromagnetic field. Over the years many attempts have been made to alleviate a significant drawback of the original theory, namely that it gives rise to unphysical, exponentially growing solutions of the electron's velocity in the absence of an externally applied force. A comprehensive review of the subject for charged particles in harmonically varying fields in classical, quantum, relativistic and non-relativistic domains may be found in [4] and references therein. An additional controversy over radiation emission when the charged particle is placed in a constant, uniform accelerating field was resolved after some debate [5]. That notwithstanding, the question has still not been fully resolved, and an equation of motion that accurately describes the motion of a radiating charge is still not available.

From a classical standpoint the electron is generally described as a point particle with rest mass $m_0$. The essence of the problem begins with the original derivation summarized in the Abraham-Lorentz equation of motion, which takes the form [4-7]:

$$m_0 \dot{\mathbf{v}} - \frac{2e^2}{3c^3} \ddot{\mathbf{v}} = \mathbf{F}_{ext}, \tag{1}$$

where the dots indicate full time derivatives, and $\mathbf{F}_{rad} = \frac{2e^2}{3c^3} \ddot{\mathbf{v}}$, the reaction force, is assumed to be a small correction compared to the externally applied force, $\mathbf{F}_{ext}$. The radiated power emitted by the accelerating electron is $P = \frac{2e^2 |\dot{\mathbf{v}}|^2}{3c^3}$, where $e$ is its charge and $c$ is the speed of light. The physical implications of the derivative of the acceleration in Eq.(1) indicates a force that changes as a function of time. The interested reader may consult reference [8] for further discussion on that topic.

The inadequacy of Eq.(1) may be ascertained by setting $\mathbf{F}_{ext} = 0$, an action that leads to a possible solution having the form $\mathbf{v} \sim \mathbf{v}_0 e^{t/\tau_0}$, where $\tau_0 = \frac{2e^2}{3m_0 c^3} \approx 3 \times 10^{-24}$ sec. Derivations, extended discussion, and range of validity of Eq.(1) may be found in references [4-7]. A method alternative to Eq.(1) that avoids unphysical, runaway solutions may be obtained by assuming the electron has uniform charge density over its volume, and thus finite radius, by defining the reaction force in terms of the applied force as follows [9,10]:

$$m_0 \frac{d\mathbf{v}}{dt} = \mathbf{F}_{ext} + \mathbf{F}_{rad} = \mathbf{F}_{ext} + \tau_0 \frac{d\mathbf{F}_{ext}}{dt}, \tag{2}$$

where $c\tau_0 \sim 10^{-13} cm$ is the classical electron radius. Although Eq.(2) had originally been presented as a Taylor series expansion [9], the authors in [10] showed that Eq.(2) is exact and valid in the quantum domain, provided a proper interpretations of the variables is made. Our present goal is to use Eq.(2) as the starting point to derive a new hydrodynamic equation of motion that includes linear and nonlinear damping contributions, and we contrast the result that emerges from our development with the hydrodynamic equation that follows from the usual introduction of a phenomenological damping coefficient [11,12], independent of the type of externally applied force. In addition, we will explore the consequences of using this approach to the dynamics that ensues in media whose physical description requires multiple polarization components.

We first develop the equation of motion of a charged particle and derive the electron polarization component for a nonrelativistic electron gas. Assuming the particle is acted upon by an external electromagnetic field that induces a Lorentz force, and that it is subject to internal electron gas pressure, Eq.2 becomes:

$$m_0 \left( \frac{\partial \mathbf{v}}{\partial t} + (\mathbf{v} \bullet \nabla) \mathbf{v} \right) = \left( e\mathbf{E} + \frac{e}{c} \mathbf{v} \times \mathbf{H} - \frac{\nabla p}{n} \right) \\ + \tau_0 \left[ \frac{\partial}{\partial t} \left( e\mathbf{E} + \frac{e}{c} \mathbf{v} \times \mathbf{H} - \frac{\nabla p}{n} \right) + (\mathbf{v} \bullet \nabla) \left( e\mathbf{E} + \frac{e}{c} \mathbf{v} \times \mathbf{H} - \frac{\nabla p}{n} \right) \right], \quad (3)$$

where $-\frac{\nabla p}{n}$ is the pressure term, and we have used the total derivative identity $\frac{d}{dt} = \frac{\partial}{\partial t} + (\mathbf{v} \bullet \nabla)$ on both sides of Eq.(2). An immediate observation that can be made in examining Eq.(3) is that the number of nonlinear source terms has more than tripled, a result that may be consequential at ultra-high intensities and/or in a relativistic context. After we identify the current density as $\mathbf{J} = ne\mathbf{v}$, where $n=n(\mathbf{r},t)$, and its partial derivative with respect to time as $\dot{\mathbf{J}} = \frac{\partial \mathbf{J}}{\partial t} = ne \frac{\partial \mathbf{v}}{\partial t} + e\mathbf{v} \frac{\partial n}{\partial t}$, Eq.(3) can be reformulated in straightforward fashion and becomes:

$$\dot{\mathbf{J}} - \frac{\dot{n}}{n} \mathbf{J} + (\mathbf{J} \bullet \nabla) \frac{\mathbf{J}}{ne} = \frac{ne^2}{m_0} \mathbf{E} + \frac{e}{m_0 c} \mathbf{J} \times \mathbf{H} - \frac{e}{m_0} \nabla p \\ + \frac{\tau_0}{m_0} \left[ \begin{array}{l} ne^2 \frac{\partial \mathbf{E}}{\partial t} + \frac{e}{c} \left( \dot{\mathbf{J}} - \frac{\dot{n}}{n} \mathbf{J} \right) \times \mathbf{H} + \frac{e}{c} \mathbf{J} \times \frac{\partial \mathbf{H}}{\partial t} - e\nabla \frac{\partial p}{\partial t} + \frac{e\nabla p}{n} \frac{\partial n}{\partial t} \\ + e(\mathbf{J} \bullet \nabla)\mathbf{E} + \frac{e}{c} (\mathbf{J} \bullet \nabla) \left( \frac{\mathbf{J}}{ne} \times \mathbf{H} \right) - (\mathbf{J} \bullet \nabla) \frac{\nabla p}{n} \end{array} \right]. \quad (4)$$

All pressure terms on the right hand side may be reworked by exploiting the continuity equation, $\dot{\rho} = e\dot{n}(\mathbf{r},t) = -\nabla \bullet \mathbf{J} = -\nabla \bullet \dot{\mathbf{P}}$, from which the solution $n = n_0 - \frac{1}{e} \nabla \bullet \mathbf{P}$ follows. $n_0$ is the charge density in the absence of applied fields, and is presently taken to be uniform throughout the volume. Assuming either an ideal or a quantum electron gas leads to the same lowest order, linear pressure contributions [12,13], which may be written for a quantum gas as follows:

$$-\frac{e\nabla p}{m_0} - \tau_0 \frac{e}{m_0}\nabla\frac{\partial p}{\partial t} + \tau_0 \frac{\dot{n}}{n}\frac{e}{m_0}\nabla p \approx \frac{5E_F}{3m_0}\nabla(\nabla\bullet\mathbf{P}) + \tau_0 \frac{5E_F}{3m_0}\nabla(\nabla\bullet\dot{\mathbf{P}}). \tag{5}$$

The last pressure term on the left hand side is purely nonlinear and its contribution is neglected in the development below. Retaining lowest order terms on both sides of the Eq.(4) we obtain:

$$\ddot{\mathbf{P}} + \frac{1}{n_0 e}\left[\dot{\mathbf{P}}\nabla\bullet\dot{\mathbf{P}} + (\dot{\mathbf{P}}\bullet\nabla)\dot{\mathbf{P}}\right] = \frac{e^2}{m_0}\left(n_0 - \frac{1}{e}\nabla\bullet\mathbf{P}\right)\mathbf{E} + \frac{e}{m_0 c}\dot{\mathbf{P}}\times\mathbf{H}$$
$$+ \frac{5E_F}{3m_0}\nabla(\nabla\bullet\mathbf{P}) + \frac{\tau_0 e^2}{m_0}\left(n_0 - \frac{1}{e}\nabla\bullet\mathbf{P}\right)\frac{\partial\mathbf{E}}{\partial t} + \tau_0 \frac{5E_F}{3m_0}\nabla(\nabla\bullet\dot{\mathbf{P}}) \tag{6}$$

The surviving terms are the only linear terms proportional to $\tau_0$ in Eq.3. Finally, using the Ampere-Maxwell's equation:

$$\left(\nabla\times\mathbf{H} - \frac{4\pi}{c}\frac{\partial\mathbf{P}}{\partial t}\right) = \frac{1}{c}\frac{\partial\mathbf{E}}{\partial t} \tag{7}$$

to substitute for $\frac{\partial\mathbf{E}}{\partial t}$, we can rewrite Eq.(6) in a manner that allows us to clearly identify linear and nonlinear damping terms:

$$\ddot{\mathbf{P}} + \omega_p^2\tau_0\left(1 - \frac{\nabla\bullet\mathbf{P}}{n_0 e}\right)\dot{\mathbf{P}} = \frac{\omega_p^2}{4\pi}\left[\mathbf{E} + c\tau_0\nabla\times\mathbf{H}\right] - \frac{e}{m_0}\nabla\bullet\mathbf{P}\left[\mathbf{E} + c\tau_0\nabla\times\mathbf{H}\right]$$
$$+ \frac{5E_F}{3m_0}\nabla(\nabla\bullet\mathbf{P}) + \tau_0\frac{5E_F}{3m_0}\nabla(\nabla\bullet\dot{\mathbf{P}}) + \frac{e}{m_0 c}\dot{\mathbf{P}}\times\mathbf{H} + \frac{1}{n_0 e}\left[\dot{\mathbf{P}}\nabla\bullet\dot{\mathbf{P}} + (\dot{\mathbf{P}}\bullet\nabla)\dot{\mathbf{P}}\right] \tag{8}$$

where $\omega_p^2 = \frac{4\pi n_0 e^2}{m_0}$ is the plasma frequency. Eq.(8) should be compared with Eq.(9) below, which can be derived under conditions similar to Eq.(2), but with the mere insertion of a phenomenological damping term, $\gamma m_0 \mathbf{v}$, to fit the experimental data relative to a chosen metal or free electron gas system [13]:

$$\ddot{\mathbf{P}} + \gamma\dot{\mathbf{P}} = \frac{n_0 e^2}{m_0}\mathbf{E} - \frac{e}{m_0}(\nabla\bullet\mathbf{P})\mathbf{E} + \frac{5E_F}{3m_0}\nabla(\nabla\bullet\mathbf{P}) + \frac{e}{m_0 c}\dot{\mathbf{P}}\times\mathbf{H} + \frac{1}{n_0 e}\left[\dot{\mathbf{P}}\nabla\bullet\dot{\mathbf{P}} + (\dot{\mathbf{P}}\bullet\nabla)\dot{\mathbf{P}}\right], \tag{9}$$

For silver or gold, for instance, the data suggests $\gamma \approx 10^{14}\sec^{-1}$, which together with a plasma wavelength of $\lambda_p \approx 300\text{nm}$, leads to an effective $\tau_{0,Ag,Au} \approx 10^{-18}\sec$ at low intensities. The electron interaction with the lattice thus increases the characteristic time by approximately six orders of magnitude compared to the equivalent $\tau_0 = \frac{2e^2}{3m_0 c^3}$ for purely free electrons, with a characteristic length that approaches the size of the atom. We will return to discuss this issue below.

We now take a closer look at Eq.(8), where we identify the damping coefficient on the left hand side of the equation, $\gamma = \omega_p^2 \tau_0 \left(1 - \frac{\nabla \cdot \mathbf{P}}{n_0 e}\right)$. In this case, $\gamma$: (i) is not the mere constant found in Eq.(9); (ii) depends on the fractional charge buildup that may occur at or near surfaces, corners or sharp edges; and (iii) becomes a nonlinear source term. Additional damping terms appear in Eq.(8) that are proportional to $\tau_0$. Using Eq.(9) it can be shown, using the identity $\frac{1}{n_0 e} \nabla \cdot \mathbf{P} = -\frac{n - n_0}{n_0}$, that the relative polarization charge density can be of order 1% at the surface of nanowires of circular cross section, for intensities of order 1GW/cm$^2$ [14], an indication that this term may impact the dynamics within hot spots generated by sharp-edged objects. Thus these effects may become important particularly in nanoscale optical geometries. Another important observation in Eq.(8) is that magnetic effects emerge as a combination of linear and nonlinear terms as perturbations to the magnitude of the electric field. It is important to note that linear, nonlocal behavior arises independent of electron gas pressure in the form of a spatial derivative of the magnetic field. However, notwithstanding the fact that for typical bulk metals $c\tau_{0,metal} \approx 10^{-7}$cm, the term $|\nabla \times \mathbf{H}| \sim kH$ could make up some of the discrepancy between the two terms and become important if: (i) large-intensity evanescent fields are excited; (ii) high local magnetic fields or currents can be generated by engineering magnetic resonances; (iii) field delocalization may occur in cavities of all types such that for the local fields $|\mathbf{E}| << |\mathbf{H}|$; or (iv) via a combination of (i)-(iii). Finally, linear, nonlocal contributions to electron gas pressure are also present, which may become important for high harmonic generation (beyond blue/violet region of the spectrum) or for ultrashort pulses.

Additional consequences from Eq.(8) may be garnered in the linear regime. So far we have considered only the response of free electrons. Consider the response of typical metals over a spectral range where it is necessary to incorporate multiple polarization components, in the form of the free electron contribution discussed above, and bound electrons as represented by Lorentz oscillators [13-15]. The analysis that follows thus applies to dielectric materials as well. We will show that the equations of motion of each polarization component is coupled to the other polarizations as a direct consequence of the partial time derivative of the electric field. Bound electrons obey an equation similar to Eq.(2). Assuming a linear restoring force $-k\mathbf{r}$ along with an applied electromagnetic field that gives rise to a Lorentz force, $e\mathbf{E}+\frac{e}{c}\dot{\mathbf{r}}\times\mathbf{H}$, we may write:

$$m_b \frac{d^2\mathbf{r}}{dt^2} = \left(-k\mathbf{r}+e\mathbf{E}+\frac{e}{c}\dot{\mathbf{r}}\times\mathbf{H}\right) + \tau_{0,b}\left[\begin{array}{l}\left(-k\dot{\mathbf{r}}+e\frac{\partial\mathbf{E}}{\partial t}+\frac{e}{c}\ddot{\mathbf{r}}\times\mathbf{H}+\frac{e}{c}\dot{\mathbf{r}}\times\dot{\mathbf{H}}\right)\\ +(\mathbf{v}\bullet\nabla)\left(-k\mathbf{r}+e\mathbf{E}+\frac{e}{c}\dot{\mathbf{r}}\times\mathbf{H}\right)\end{array}\right] \quad , \quad (10)$$

where $k$ is the spring constant, and the subscript $b$ helps us differentiate between free and bound effective masses and characteristic times. Neglecting all nonlinear terms in both Eqs.(8) and (10) we determine the lowest order, linear contributions of damping and any interplay that may ensue between free and bound electrons. Multiplying Eq.(10) by $n_b e$, where $n_b$ is now assumed to be constant for bound charges, and keeping only linear terms we obtain:

$$\ddot{\mathbf{P}}_b + \tau_{0,b}\omega_{0,b}^2\dot{\mathbf{P}}_b = -\omega_{0,b}^2\mathbf{P}_b + \frac{n_b e^2}{m_b}\mathbf{E} + \frac{\tau_{0,b}n_b e^2}{m_b}\frac{\partial\mathbf{E}}{\partial t} \quad , \quad (11)$$

where $\mathbf{P}_b = n_b e\mathbf{r}$; $\omega_{0,b}^2 = \frac{k}{m_b}$ and $\omega_{p,b}^2 = \frac{4\pi n_b e^2}{m_b}$ are the resonance and plasma frequencies, respectively, of the bound electron system. An equation similar to Eq.(11) is also discussed in reference [6] for a single polarization component, but without the development that follows. For multiple polarization species Ampere-Maxwell's equation takes the form:

$$\nabla\times\mathbf{H} - \frac{4\pi}{c}\frac{\partial\mathbf{P}_{Total}}{\partial t} = \frac{1}{c}\frac{\partial\mathbf{E}}{\partial t} \quad , \quad (12)$$

where the total polarization is the vector sum of the two components, $\mathbf{P}_{Total} = \mathbf{P}_b + \mathbf{P}_f$. We have adopted the subscript $f$ to describe the free electron polarization. Combining Eqs.(8), (11) and (12) leads to the following coupled, linear equations of motion:

$$\ddot{\mathbf{P}}_b + \tau_{0,b}\left(\omega_{0,b}^2+\omega_{p,b}^2\right)\dot{\mathbf{P}}_b + \tau_{0,b}\omega_{p,b}^2\dot{\mathbf{P}}_f + \omega_{0,b}^2\mathbf{P}_b = \frac{n_b e^2}{m_b}\left(\mathbf{E}+c\tau_{0,b}\nabla\times\mathbf{H}\right)$$

$$\ddot{\mathbf{P}}_f + \tau_0\omega_p^2\dot{\mathbf{P}}_f + \tau_0\omega_p^2\dot{\mathbf{P}}_b = \frac{n_0 e^2}{m_0}\left(\mathbf{E}+c\tau_0\nabla\times\mathbf{H}\right)$$

. (13)

The introduction of damping via Eq.(2) thus appears to alter the linear dynamics by directly coupling the two polarizations of the system. Clearly this is not the case in any traditional analysis, where one simply injects a phenomenological damping coefficient into each of Eqs.(13) [13-15], a procedure that conceals coupling and magnetic effects. We also note that, similarly to what occurs in the free electron portion of the material, damping introduces a nonlocal contribution in the dynamics of bound charges as well, in the form of a spatial derivative of the magnetic field. An examination of the data of noble metals like silver and gold in the visible and UV ranges suggests that bound oscillators (d-shell electrons) may be damped at a rate $\gamma_b \approx 10^{15} \sec^{-1}$ [16], corresponding to $\tau_{0,b} \approx 10^{-17} \sec$. As a result, in the absence of magnetic resonances or surface currents that could enhance the magnetic field or its derivative, magnetic terms may be ignored in both Eqs.(13). Since for typical metals each atom contributes one conduction electron, for simplicity we may assume that $n_b \approx n_0$; $m_b = m_0$; and that $\tau_{0,b}\left(\omega_{0,b}^2 + \omega_{p,b}^2\right) \approx 2\tau_{0,b}\omega_{p,b}^2 = 2\gamma_b$. The latter assumption implies that the resonance frequency is of the same order of magnitude as the plasma frequency. Then, taking the Fourier transform of Eqs.(13) we obtain a coupled pair of polarization equations:

$$\left(-\omega^2 - i2\gamma_b\omega + \omega_0^2\right)\mathbf{P}_b - i\gamma_b\omega\mathbf{P}_f \approx \frac{n_0 e^2}{m_0}\mathbf{E}$$
$$\left(-\omega^2 - i\gamma_f\omega\right)\mathbf{P}_f - i\gamma_f\omega\mathbf{P}_b \approx \frac{n_0 e^2}{m_0}\mathbf{E}$$

(14)

Eliminating the free electron polarization and solving for the bound and free electron polarizations, respectively, yields:

$$\mathbf{P}_b = \frac{n_0 e^2}{m_0\left(-\omega^2 - i2\gamma_b\omega + \omega_0^2\right)} \frac{\left(1 + \frac{i\gamma_b\omega}{\left(-\omega^2 - i\gamma_f\omega\right)}\right)}{\left(1 + \frac{\gamma_b\gamma_f\omega^2}{\left(-\omega^2 - i\gamma_f\omega\right)\left(-\omega^2 - i2\gamma_b\omega + \omega_0^2\right)}\right)} \mathbf{E},$$

(15)

and

$$\mathbf{P}_f = \frac{n_0 e^2}{m_0\left(-\omega^2 - i\gamma\omega\right)} \frac{\left(1 + \frac{i\gamma_f\omega}{\left(-\omega^2 - i2\gamma_b\omega + \omega_0^2\right)}\right)}{\left(1 + \frac{\gamma_b\gamma_f\omega^2}{\left(-\omega^2 - i\gamma_f\omega\right)\left(-\omega^2 - i2\gamma_b\omega + \omega_0^2\right)}\right)} \mathbf{E}.$$

(16)

The bound electron polarization is explicitly influenced by the dynamics of free electrons, and vice versa. In this regard we note that at high frequencies metals are typically characterized by many bound electron resonances [16]. Also at high frequencies the number of (d-shell) electrons each atom contributes to the dielectric constant increases [17], boosting the effective bound electron plasma frequency. As a result, the $\mathbf{P}_b$ in the second of Eqs.(14) turns into a sum over all possible resonances that simultaneously contribute to the damping of free electrons in a manner similar to Eqs.(15-16), i.e. $\mathbf{P}_{bound} = \sum_{j=1}^{N} \mathbf{P}_j(\omega_{0,j}, \omega_{p,j})$, where $\omega_{0,j}$ and $\omega_{p,j}$ are, respectively, the resonance and plasma frequencies associated with the *jth* oscillator. By the same token, each bound oscillator feels the effects of all other polarization components, including that of free electrons. These circumstances, and the presence of the lattice together add viscous terms that account for modes of decay other than photon emission that help to partially explain why the observed damping coefficients and characteristic times associated with nearly-free electrons in metals are very different compared to their corresponding nominal values for purely free electrons. Adding Eqs.(15) and (16) yields a suggestive expression for the total polarization:

$$\mathbf{P}_{total} = \left[ \frac{n_0 e^2}{m_0(-\omega^2 - i\gamma_f \omega)} + \frac{n_0 e^2}{m_0(-\omega^2 - i2\gamma_b \omega + \omega_0^2)} \left( \frac{1 + \frac{i(\gamma_b + \gamma_f)\omega}{(-\omega^2 - i\gamma_f \omega)} - \frac{\gamma_f \gamma_b \omega^2}{(-\omega^2 - i\gamma_f \omega)^2}}{1 + \frac{\gamma_f \gamma_b \omega^2}{(-\omega^2 - i\gamma_f \omega)(-\omega^2 - i2\gamma_b \omega + \omega_0^2)}} \right) \right] \mathbf{E}, \quad (17)$$

from which a dielectric constant for a two-component medium may be determined:

$$\varepsilon(\omega) = 1 - \frac{4\pi n_0 e^2}{m_0(-\omega^2 - i\gamma_f \omega)} + \frac{4\pi n_0 e^2}{m_0(-\omega^2 - i2\gamma_b \omega + \omega_0^2)} \left( \frac{1 + \frac{i(\gamma_b + \gamma_f)\omega}{(-\omega^2 - i\gamma_f \omega)} - \frac{\gamma_f \gamma_b \omega^2}{(-\omega^2 - i\gamma_f \omega)^2}}{1 + \frac{\gamma_f \gamma_b \omega^2}{(-\omega^2 - i\gamma_f \omega)(-\omega^2 - i2\gamma_b \omega + \omega_0^2)}} \right). \quad (18)$$

While the free electron polarization dominates away from the bound electron resonance, we note that in practical terms free and bound polarizations are no longer individually distinguishable, because both polarizations are intertwined, as Eqs.(15-16) suggest. For a given set of parameters the extra factor that multiplies the second term in Eq.(18) clearly modifies the dielectric constant of the uncoupled system. In Fig.1 we plot the complex $\varepsilon(\omega)$ of a hypothetical gold-like metal (i.e. with $\gamma_b$ and $\gamma_f$ that are roughly those used to fit gold data, as described above) with and without the extra factor in Eq.(18): the absorption resonance redshifts by approximately 30nm, while amplitude changes are evident in both the real and imaginary parts. The changes for the imaginary dielectric function are especially large at longer wavelengths. Applying our alternative formalism, integration of either Eqs.(13) and/or their fully nonlinear counterparts should proceed by fitting the measured dielectric constant by first retrieving oscillator parameters using the full Eq.(18) if, for example, a two-component medium is used, rather than fitting Eq.(18) without the extra factor, as is usually done [13-15].

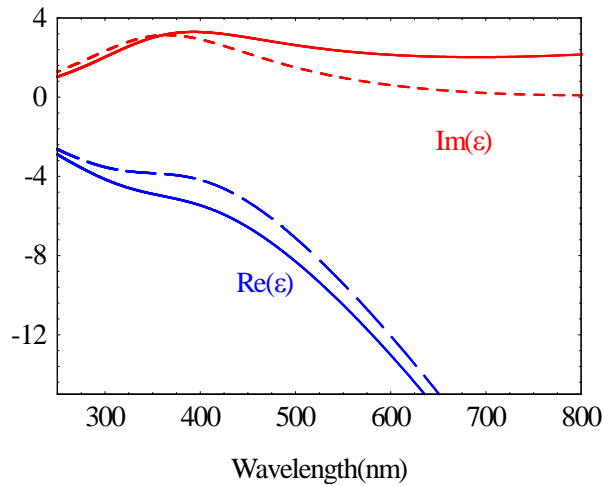

**Fig.1**: Complex dielectric constant plotted with (dashed lines) and without (solid lines) the extra factor in the second term of Eq.(17) for a gold-like material modeled by one Drude and one Lorentz oscillator with resonance frequency near 400nm.

In summary, we have applied a new formalism starting with the equation of motion with radiation reaction/damping to derive a modified hydrodynamic equation where the damping term is a function of local charge density. The formalism also introduces new linear, nonlinear, and nonlocal source terms that depend on the spatial derivative of the magnetic field, which under the right conditions can influence both linear and nonlinear dynamics, e.g. circular dichroism and harmonic generation. We also showed that in a multi-component medium, well-represented by a typical metal at near-IR wavelengths and below, or by systems where both free electrons and ionic species are the result of ionization, as may occur at high intensities or in a relativistic regime [18], the new, more general formulation generates interference between the polarizations, triggering shifts in the position of the resonances and changes in the amplitude of the dielectric constant, a fact that requires a reassessment of the oscillator parameters that are used to fit any measured data. Of some interest will be the development of Eq.(2) in the relativistic regime [19].

**Acknowledgment**

Financial support from U.S. Army RDECOM Acquisition Grant No. W911NF-15-1-0178 for JWH is gratefully acknowledged. JT and CC acknowledge financial support from RDECOM Grant W911NF-16-1-0563 from the International Technology Center-Atlantic.

**References**

1. H. A. Lorentz, *The Theory of Electrons and its Applications to the Phenomena of Light and Radiant Heat* (Columbia University Press, New York 1909). The second edition appeared in 1915 and was reprinted by Dover.
2. M. Abraham and A. Föppl, *Theorie der Elektrizitat*, Vol. II (Teubner, Leipzig, 1905).
3. P. A. M. Dirac, "Classical Theory of Radiating Electrons," Proc. Roy. Soc. A **167**, 148 (1938).
4. F. Rohrlich, "The dynamics of a charged sphere and the electron," Am. J. Phys. **65**, 1051-1056 (1997).
5. See, for instance: R. Hammond, "Charged Particle in a Constant, Uniform Electric Field with Radiation Reaction," Adv. Studies Theor. Phys. **5**, 275-282 (2011) and references therein.
6. J. D. Jackson, "*The Classical Electromagnetic field,*" (Wiley, 1999).
7. M. A. Heald and J. B. Marion, *Classical Electromagnetic Radiation*, (Saunders, 1995).
8. D. Eager, A.-M. Pendrill, N. Reistad, "Beyond velocity and acceleration: jerk, snap and higher derivatives," *Eur. J. Phys.* **37** 065008 (2016).
9. E.J. Moniz and D.H. Sharp, "Radiation reaction in nonrelativistic quantum electrodynamics," Phys. Rev. D **15,** 2850-2865 (1977).
10. G.W. Ford and R.F. O'Connell, "Radiation reaction in electrodynamics and the elimination of runaway solutions," Phys. Lett. A. **157**, 217-220 (1991); ibid, "Alternative equations of motion for the radiating electron," Appl. Phys. B **60**, 301-302 (1995).
11. N. Bloembergen and Y. R. Shen, "Optical nonlinearity of a plasma," Phys. Rev. **141**, 298-305 (1966).
12. J. E. Sipe, V. C. Y. So, M. Fukui and G. I. Stegeman, "Analysis of second-harmonic generation at metal surfaces", Phys. Rev. B **21**, 4389-4402 (1980).
13. M. Scalora, M. A. Vincenti, D. de Ceglia, V. Roppo, M. Centini, N. Akozbek and M. J. Bloemer, "Second- and third-harmonic generation in metal-based structures," Phys. Rev. A **82**, 043828 (2010).

14. M. Scalora, D. de Ceglia, M. A. Vincenti and J. W. Haus, "Nonlocal and Quantum Tunneling Contributions to Harmonic Generation in Nanostructures: Electron Cloud Screening Effects", Phys. Rev. A **90**, 013831 (2014).
15. M. Scalora, M. A. Vincenti, D. de Ceglia, M. Grande, and J. W. Haus, "Raman scattering near metal nanostructures," J. Opt. Soc. Am. B **29**, 2035-2045, (2012); ibid, J. Opt. Soc. Am. B **30**, 2634-2639 (2013).
16. A. D. Rakić, A. B. Djurišić, J. M. Elazar, and M. L. Majewski, "Optical properties of metallic films for vertical-cavity optoelectronic devices," Appl. Optics **37**, 5271 (1998).
17. H. Ehrenreich and H. R. Philipp, "Optical properties of Ag and Cu", Phys. Rev. **128**, 1622 (1962).
18. E. S. Sarachick and G. T. Schappert, "Classical theory of the scattering of intense laser radiation by free electrons," Phys. Rev. D **1** 2738 (1970).
19. G.W. Ford and R.F. O'Connell, "Relativistic form of radiation reaction," Phys. Lett. A **174,** 182 (1993).